\documentclass[prb,showpacs,twocolumn,nofloats,endfloats]{revtex4}

\usepackage{amsmath,amssymb,epsfig,epsf}

\begin{document}

\title
{\bf Mott--Hubbard Transition and Anderson Localization: 
Generalized Dynamical Mean-Field Theory Approach}
\author{E.Z. Kuchinskii, I.A. Nekrasov, M.V. Sadovskii}

\affiliation
{Institute for Electrophysics, Russian Academy of Sciences,
Ekaterinburg, 620016, Russia}

\begin{abstract}

Density of states, dynamic (optical) conductivity and phase diagram of 
strongly correlated and strongly disordered paramagnetic Anderson--Hubbard 
model are analyzed within the generalized dynamical mean field theory 
(DMFT+$\Sigma$ approximation).  Strong correlations are accounted by DMFT, 
while disorder is taken into account via the appropriate generalization of 
self-consistent theory of localization.  The DMFT effective single impurity 
problem is solved by numerical renormalization group (NRG) and we consider 
the three-dimensional system with semi-elliptic density of states.  
Correlated metal, Mott insulator and correlated Anderson insulator phases are
identified via the evolution of density of states and dynamic conductivity,
demonstrating both Mott-Hubbard and Anderson metal-insulator transition and
allowing the construction of complete zero-temperature phase diagram of
Anderson--Hubbard model. Rather unusual is the possibility of disorder 
induced Mott insulator to metal transition.

\end{abstract}


\maketitle

\newpage

\section{Introduction}

The importance of the electronic interaction and randomness for the properties 
of condensed matter is well known \cite{Lee85}. 
Coulomb correlations and disorder are both driving forces of
metal--insulator transitions (MITs) connected with the localization and 
delocalization of particles. In particular, the Mott--Hubbard MIT is caused
by the electronic repulsion \cite{Mott90}, while Anderson MIT is 
due to random scattering of non--interacting particles \cite{Anderson58}. 
Actually, disorder and interaction effects are known to compete in many subtle
ways \cite{Lee85,ma} and this problem becomes much more complicated in the case
of strong electron correlations and strong disorder, determining the physical
mechanisms of Mott--Anderson MIT \cite{Lee85}. 

The cornerstone of the modern theory of strongly correlated systems is 
the dynamical mean--field theory (DMFT) 
\cite{MetzVoll89,vollha93,pruschke,georges96}, constituting 
a non--perturbative theoretical framework for the investigation of correlated
lattice electrons with a local interaction.
In this approach the effect of local disorder can be taken into account
through the standard average density of states (DOS) \cite{ulmke95} 
in the absence of interactions, leading to the well known coherent potential 
approximation \cite{vlaming92}, which does not describe the physics of  
Anderson localization. To overcome this deficiency Dobrosavljevi\'{c} and 
Kotliar \cite{Dobrosavljevic97} formulated a variant of the DMFT where the
geometrically averaged local DOS was computed from the solutions of the
self--consistent stochastic DMFT equations. 
Subsequently, Dobrosavljevi\'{c} \emph{et al.} \cite{Dobrosavljevic03} 
incorporated the geometrically averaged local DOS into the self--consistency 
cycle and derived  a mean--field theory of Anderson localization which
reproduced many of the expected features of the disorder--driven
MIT for non--interacting electrons. This approach was extended by Byczuk 
\emph{et al.} \cite{BV} to include Hubbard correlations via DMFT, which lead
to highly non trivial phase diagram of Anderson--Hubbard model with correlated
metal, Mott insulator and correlated Anderson insulator phases. 
The main deficiency of these approaches, however, is inability to calculate 
directly measurable physical properties, such as conductivity, which is of 
major importance and defines MIT itself.

At the same time the well developed approach of self--consistent theory of
Anderson localization, based on solving the equations for the
generalized diffusion coefficient, demonstrated its efficiency in 
non--interacting case long ago \cite{VW,WV,MS,MS86,VW92,Diagr} and some attempts 
to include interaction effects within this approach were also undertaken with
some promising results \cite{MS86,KSS}. However, up to now there were no 
attempts to incorporate this approach to the modern theory of strongly 
correlated electronic systems. Here we undertake such research, studying both
Mott--Hubbard and Anderson MITs via direct calculations of both the average
DOS and dynamic (optical) conductivity. 

Our approach is based on recently proposed generalized DMFT+$\Sigma$
approximation \cite{JTL05,PRB05,FNT06,PRB07}, which
on the one hand retains the single-impurity description of the DMFT, 
with a proper account for local Hubbard--like correlations and the 
possibility to use impurity solvers like NRG\cite{NRG,BPH,Bull}, while on the 
other hand, allows one to include additional (either local or non-local) 
interactions (fluctuations) on a non-perturbative model basis.

Within this approach we have already studied both single- and two-particle 
properties, of two--dimensional Hubbard model, concentrating mainly on 
the problem of pseudogap formation in the density of states of the 
quasiparticle band in both correlated metals and doped Mott insulators, 
with an application to superconducting cuprates. We analyzed the
evolution of non--Fermi liquid like spectral density and ARPES spectra 
\cite{PRB05}, ``destruction'' of Fermi surfaces and formation of Fermi ``arcs'' 
\cite{JTL05}, as well as pseudogap anomalies of optical conductivity 
\cite{PRB07}. 
Briefly we also considered impurity scattering effects \cite{FNT06}. 

In this paper we apply our DMFT+$\Sigma$ approach for 
calculations of the density of states, dynamic conductivity and phase 
diagram of strongly correlated and strongly disordered three--dimensional
paramagnetic Anderson--Hubbard model. 
Strong correlations are again accounted by DMFT, 
while disorder is taken into account via the appropriate generalization of 
self-consistent theory of localization. 

The paper is organized as follows:
In section \ref{leng_intro} we present a short description of our generalized
DMFT+$\Sigma$ approximation with application to disordered Hubbard model. 
In section \ref{opt_cond} we present basic
DMFT+$\Sigma$ expressions for dynamic (optical) conductivity and
formulate appropriate self--consistent equations for the
generalized diffusion coefficient.
Computational details and results for density of states and dynamic 
conductivity are given in section \ref{results}, where we also analyze the
phase diagram of strongly disordered Hubbard model, following from our approach. 
The paper is ended with a short summary 
section~\ref{concl} including a discussion of some related problems.

\section{Basics of DMFT+$\Sigma$ approach}
\label{leng_intro}

Our aim is to consider non--magnetic disordered Anderson--Hubbard model
(mainly) at half--filling for arbitrary interaction and disorder strengths. 
Mott--Hubbard and Anderson MITs will be investigated on an equal footing. The
Hamiltonian of the model under study is written as:
\begin{equation}
H=-t\sum_{\langle ij\rangle \sigma }a_{i\sigma }^{\dagger }a_{j\sigma
}+\sum_{i\sigma }\epsilon _{i}n_{i\sigma }+U\sum_{i}n_{i\uparrow
}n_{i\downarrow },  
\label{And_Hubb}
\end{equation}
where $t>0$ is the amplitude for hopping between nearest neighbors, $U$ is
the on--site repulsion, $n_{i\sigma }=a_{i\sigma }^{\dagger }a_{i\sigma }^{{%
\phantom{\dagger}}}$ is the local electron number operator, $a_{i\sigma }$ 
($a_{i\sigma }^{\dagger}$) is the annihilation (creation) operator of an electron with spin $\sigma$,
and the local ionic energies $\epsilon _{i}$ at different lattice sites are 
considered to be independent random variables. To simplify diagrammatics in the
following we assume Gaussian probability distribution for $\epsilon _{i}$:
\begin{equation}
\mathcal{P}(\epsilon _{i})=\frac{1}{\sqrt{2\pi}\Delta}\exp\left(
-\frac{\epsilon_{i}^2}{2\Delta^2}
\right)
\label{Gauss}
\end{equation}
Here the parameter $\Delta$ is just a measure of disorder strength, and
Gaussian (``white'' noise) random field of energy level $\epsilon_i$ at lattice 
cites produces ``impurity'' scattering, leading to the standard diagram 
technique for calculation on the averaged Green's functions \cite{Diagr}.

\begin{figure}
\includegraphics[clip=true,width=0.3\textwidth]{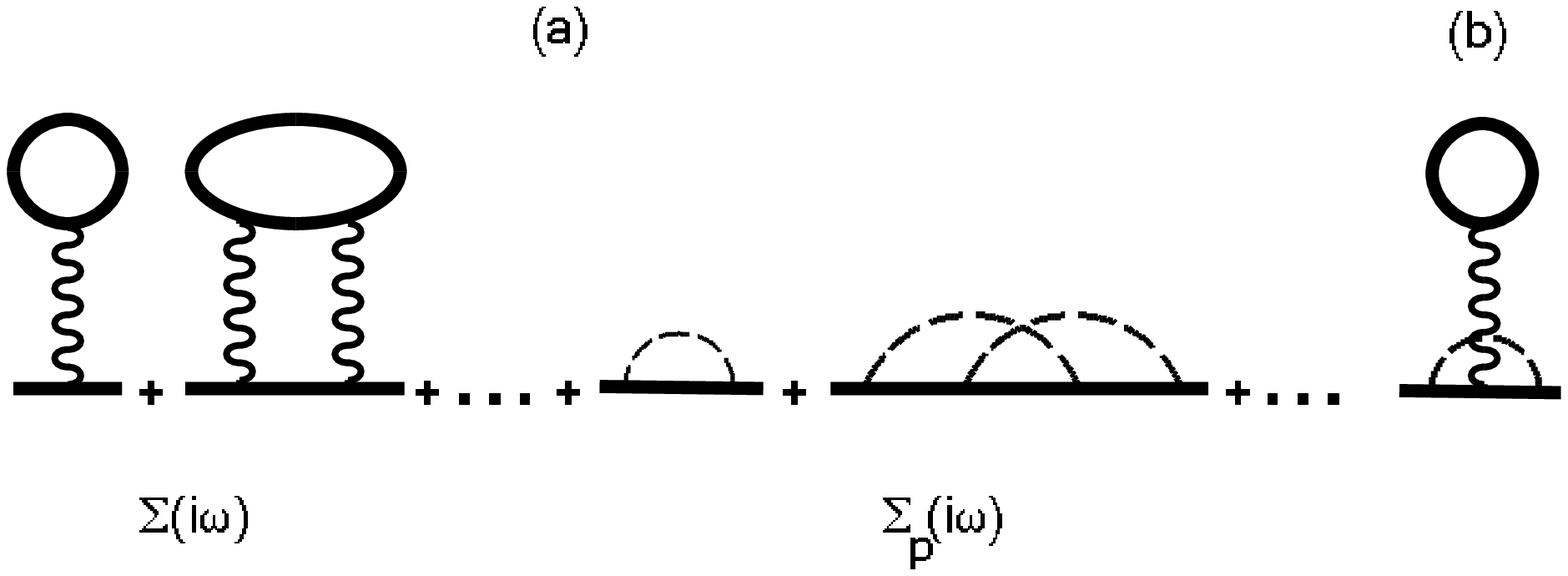}
\caption{Typical ``skeleton'' diagrams for the self--energy in the
DMFT+$\Sigma$ approach.
The first two terms are examples of DMFT self--energy diagrams; the
middle two diagrams show contributions due to random impurity scattering
represented as dashed lines. The last diagram (b) is an example of neglected 
diagram leading to interference between the local Hubbard interaction and
impurity scattering.}
\label{dDMFT_PG}
\end{figure}

DMDF+$\Sigma$ approach was initially proposed \cite{JTL05,PRB05,FNT06} as a 
simple method to include non--local fluctuations, essentially of arbitrary
nature, to the standard DMFT. In fact it can be used to include into DMFT any 
additional interaction in the following way.  Working at finite
temperatures $T$ we write down Matsubara ``time'' Fourier transformed 
single-particle Green function of the Hubbard model as:
\begin{equation}
G(i\varepsilon,{\bf p})=\frac{1}{i\varepsilon+\mu-\epsilon({\bf p})-
\Sigma(i\varepsilon)
-\Sigma_{\bf p}(i\varepsilon_n)}, \varepsilon=\pi T(2n+1),
\label{Gk}
\end{equation}
where $\epsilon({\bf p})$ is the single particle spectrum, corresponding 
to free part of (\ref{And_Hubb}), $\mu$ is chemical potential fixed by
electron concentration, and
$\Sigma(i\varepsilon)$ is the local contribution to self--energy due to Hubbard
interaction, of DMFT type (surviving in the limit of spatial dimensionality 
$d\to\infty$), while $\Sigma_{\bf p}(i\varepsilon)$ is some additional
(in general momentum dependent) self--energy part. This last 
contribution can be due e.g. to electron interactions with some 
``additional'' collective modes or order parameter fluctuations within the
Hubbard model itself. But actually it can be due to any other interactions 
(fluctuations) outside the standard Hubbard model, e.g. due to phonons or 
to random impurity scattering, 
when it is in fact local (momentum independent).  The last interaction will be
our main interest in the present paper.  Basic assumption here is the neglect 
of all interference processes of the local Hubbard interaction and 
``external'' contributions due to these additional scatterings (non-crossing 
approximation for appropriate diagrams) \cite{PRB05}, as illustrated by 
diagrams in Fig. \ref{dDMFT_PG}.

The self--consistency equations of generalized DMFT+$\Sigma$ 
approach are formulated as follows \cite{JTL05,PRB05}:
\begin{enumerate}
\item{Start with some initial guess of local self--energy
$\Sigma(i\varepsilon)$, e.g. $\Sigma(i\varepsilon)=0$}.  
\item{Construct $\Sigma_{\bf p}(i\varepsilon)$ within some (approximate) 
scheme, taking into account interactions with ``external'' interaction 
(impurity scattering in our case) which in general can depend on 
$\Sigma(i\omega)$ and $\mu$.} 
\item{Calculate the local Green function 
\begin{equation} 
G_{ii}(i\varepsilon)=\frac{1}{N}\sum_{\bf p}\frac{1}{i\varepsilon+\mu
-\epsilon({\bf p})-\Sigma(i\varepsilon)-\Sigma_{\bf p}(i\varepsilon)}.
\label{Gloc}
\end{equation}
}
\item{Define the ``Weiss field''
\begin{equation}
{\cal G}^{-1}_0(i\varepsilon)=\Sigma(i\varepsilon)+G^{-1}_{ii}(i\varepsilon).
\label{Wss}
\end{equation}
}
\item{Using some ``impurity solver'' calculate the single-particle Green 
function $G_d(i\varepsilon)$ for the effective Anderson impurity 
problem, placed at lattice site $i$ and defined by effective action
which is written, in obvious notations, as:
\begin{eqnarray}
S_{\text{eff}}=-\int_{0}^{\beta}d\tau_1\int_{0}^{\beta}
d\tau_2c_{i\sigma}(\tau_1){\cal G}^{-1}_0(\tau_1-\tau_2)c^+_{i\sigma}(\tau_2)
\nonumber\\
+\int_{0}^{\beta}d\tau Un_{i\uparrow}(\tau)n_{i\downarrow}(\tau). 
\label{Seff}
\end{eqnarray}
Actually, in the following, for the ``impurity solver'' we use NRG 
\cite{NRG,BPH,Bull}, which allows us to deal also with real frequencies, thus 
avoiding the complicated problem of analytical continuation from Matsubara 
frequencies.  
} 
\item{Define a {\em new} local self--energy \begin{equation} 
\Sigma(i\omega)={\cal G}^{-1}_0(i\omega)-G^{-1}_{d}(i\omega).
\label{StS}
\end{equation}
}
\item{Using this self--energy as 
``initial'' one in step 1, continue the procedure until (and if) convergency 
is reached to obtain 
\begin{equation} 
G_{ii}(i\varepsilon)=G_{d}(i\varepsilon).  
\label{G00}
\end{equation}
}
\end{enumerate}
Eventually, we get the desired Green function in the form of (\ref{Gk}),
where $\Sigma(i\varepsilon)$ and $\Sigma_{\bf p}(i\varepsilon)$ are those 
appearing at the end of our iteration procedure.

For $\Sigma_{\bf p}(i\varepsilon_n)$ of the random impurity problem we shall 
use the simplest possible one--loop contribution, given by the third diagram 
of Fig. \ref{dDMFT_PG} (a), neglecting ``crossing'' diagrams like the fourth 
one in Fig. \ref{dDMFT_PG} (a), i.e. just the self--consistent Born approximation 
\cite{Diagr}, which in the case of Gaussian disorder (\ref{Gauss}) leads to the 
usual expression:
\begin{equation}
\Sigma_{\bf p}(i\varepsilon)=\Delta^2\sum_{\bf p}G(i\varepsilon,{\bf p})
\equiv \Sigma_{imp}(i\varepsilon)
\label{BornSigma}
\end{equation}
which is actually ${\bf p}$-independent (local).

\section{Dynamic conductivity in DMFT+$\Sigma$ approach}
\label{opt_cond}

\subsection{Basic expressions for optical conductivity}

Physically it is clear that calculations of dynamic 
conductivity are the most direct way to study MITs, as its frequency dependence 
along with static value at zero frequency of an external field allows the 
clear distinction between metallic and insulating phases (at zero temperature
$T=0$). 

To calculate dynamic conductivity we use the general expression relating it
to retarded density--density correlation function $\chi^R(\omega,{\bf q})$
\cite{VW,Diagr}:
\begin{equation}
\sigma(\omega)=-\lim_{q\to 0}\frac{ie^2\omega}{q^2}\chi^R(\omega,{\bf q}),
\label{cond_op}
\end{equation}
where $e$ is electronic charge.

\begin{figure}
\includegraphics[clip=true,width=0.3\textwidth]{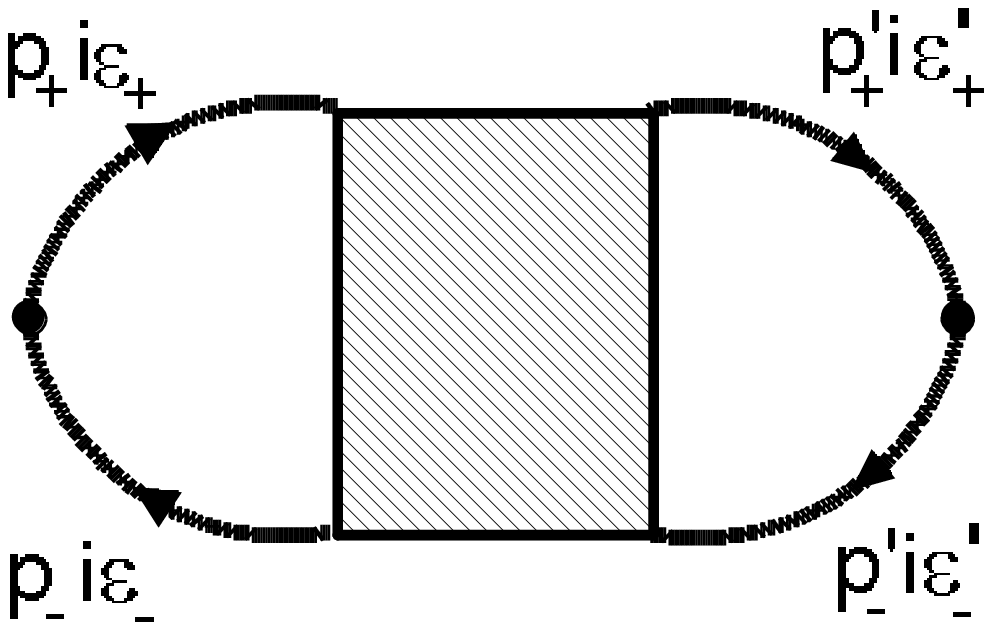}
\caption{Full polarization loop with vertex part, describing all interactions 
and impurity scatterings in particle--hole channel. The loop without vertex
corrections is included implicitly. 
Here ${\bf p}_{\pm}={\bf p}\pm\frac{\bf q}{2}$, $\varepsilon_{\pm}=
\varepsilon\pm\frac{\omega}{2}$. }
\label{loop}
\end{figure}

Next we briefly follow the derivation presented in detail in Ref. \cite{PRB07}
for the pseudogap problem, with necessary modifications for the present case.
Consider full polarization loop graph in Matsubara representation 
shown in Fig. \ref{loop}, which is conveniently  
(with explicit frequency summation) written as:
\begin{equation}
\Phi(i\omega,{\bf q})=\sum_{\varepsilon\varepsilon'}
\Phi_{i\varepsilon i\varepsilon'}(i\omega,{\bf q})\equiv\sum_{\varepsilon}
\Phi_{i\varepsilon}(i\omega,{\bf q})
\label{PhiM}
\end{equation}
and contains all possible interactions of our model, described by the full
shaded vertex part. Actually we implicitly assume here that simple loop
contribution without vertex corrections is also included in Fig. \ref{loop},
which shortens further diagrammatic expressions \cite{PRB07}.
Retarded density--density correlation 
function is determined by appropriate analytic continuation of this loop and 
can be written as\cite{VW}:
\begin{eqnarray}
\chi^R(\omega,{\bf q})=
\int_{-\infty}^{\infty}\frac{d\varepsilon}{2\pi i}\left\{\left[f(\varepsilon_+) 
-f(\varepsilon_-)\right]\Phi_{\varepsilon}^{RA}({\bf q},\omega)\right. \nonumber\\
\left.+f(\varepsilon_-)\Phi_{\varepsilon}^{RR}({\bf q},\omega)
-f(\varepsilon_+)\Phi_{\varepsilon}^{AA}({\bf q},\omega)\right\}
\label{cond_gen},
\end{eqnarray}
where $f(\varepsilon)$ -- Fermi distribution,
$\varepsilon_{\pm}=\varepsilon\pm\frac{\omega}{2}$, while two--particle loops  
$\Phi_{\varepsilon}^{RA}({\bf q},\omega)$, 
$\Phi_{\varepsilon}^{RR}({\bf q},\omega)$, 
$\Phi_{\varepsilon}^{AA}({\bf q},\omega)$ are determined by appropriate
analytic continuations $(i\varepsilon+i\omega\to\varepsilon+\omega+i\delta, 
i\varepsilon\to\varepsilon\pm i\delta, \delta\to +0)$ in (\ref{PhiM}).
Then we can write dynamic (optical) conductivity as:
\begin{eqnarray}
&&\sigma(\omega)=\lim_{q\to 0}\left(-\frac{e^2\omega}{2\pi q^2}\right)
\int_{-\infty}^{\infty}d\varepsilon\nonumber\\
&&\left\{\left[f(\varepsilon_+)
-f(\varepsilon_-)\right]\left[\Phi_{\varepsilon}^{RA}({\bf q},\omega)
-\Phi_{\varepsilon}^{RA}(0,\omega)\right]\right.\nonumber\\
&&\left.+f(\varepsilon_-)\left[\Phi_{\varepsilon}^{RR}({\bf q},\omega)
-\Phi_{\varepsilon}^{RR}(0,\omega)\right]\right.\nonumber\\
&&\left.-f(\varepsilon_+)\left[\Phi_{\varepsilon}^{AA}({\bf q},\omega)-
\Phi_{\varepsilon}^{AA}(0,\omega)\right]\right\},
\label{cond_gener}
\end{eqnarray}
where the total contribution of additional terms with zero $q$ can be shown 
(with the use of general Ward identities) to be zero. 

\begin{figure}
\includegraphics[clip=true,width=0.3\textwidth]{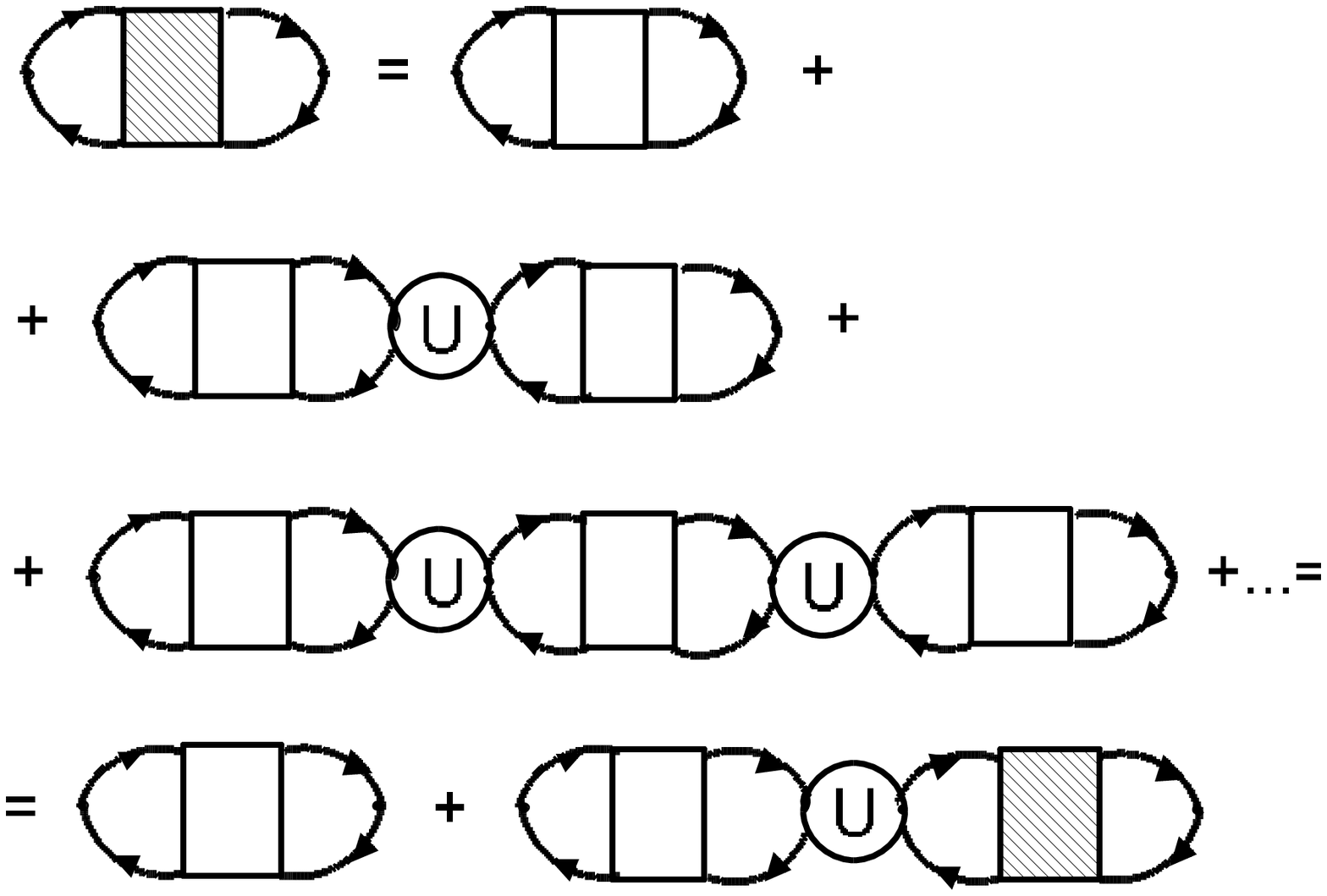}
\caption{Bethe--Salpeter equation for polarization loop in
DMFT+$\Sigma$ approach. Circle represents irreducible vertex part in 
particle--hole channel of DMFT, which contains only local Hubbard 
interactions. Unshaded rectangular vertex 
represents corrections from impurity scattering only, implicitly including the 
case free particle--hole propagation.}
\label{BS_loop}
\end{figure}

To calculate $\Phi_{i\varepsilon i\varepsilon'}(i\omega,{\bf q})$,
entering the sum over Matsubara frequencies in (\ref{PhiM}), 
in DMFT+$\Sigma$ approximation, which
neglects interference between local Hubbard interaction and impurity scatterings, 
we can write down Bethe--Salpeter equation,
shown diagrammatically in Fig. \ref{BS_loop}, where we introduce
irreducible (local) vertex $U_{i\varepsilon i\varepsilon'}(i\omega)$ 
of DMFT and ``rectangular'' vertex, containing all interactions with impurities.
Analytically this equation can be written as:
\begin{eqnarray}
&&\Phi_{i\varepsilon i\varepsilon'}(i\omega,{\bf q})=
\Phi^0_{i\varepsilon}(i\omega,{\bf q})
\delta_{\varepsilon\varepsilon'}\nonumber\\
&&+\Phi^0_{i\varepsilon}(i\omega,{\bf q})\sum_{\varepsilon''}
U_{i\varepsilon i\varepsilon''}(i\omega)
\Phi_{i\varepsilon'' i\varepsilon'}(i\omega,{\bf q}),
\label{BS_equ}
\end{eqnarray}
where $\Phi^0_{i\varepsilon}(i\omega,{\bf q})$ is the desired
function calculated neglecting vertex corrections due to Hubbard interaction 
(but taking into account all interactions due to impurity scattering).
Note that all $q$-dependence here is determined by 
$\Phi^0_{i\varepsilon}(i\omega,{\bf q})$ as the vertex 
$U_{i\varepsilon i\varepsilon'}(i\omega)$ is local and $q$-independent. 

\begin{figure}
\includegraphics[clip=true,width=0.4\textwidth]{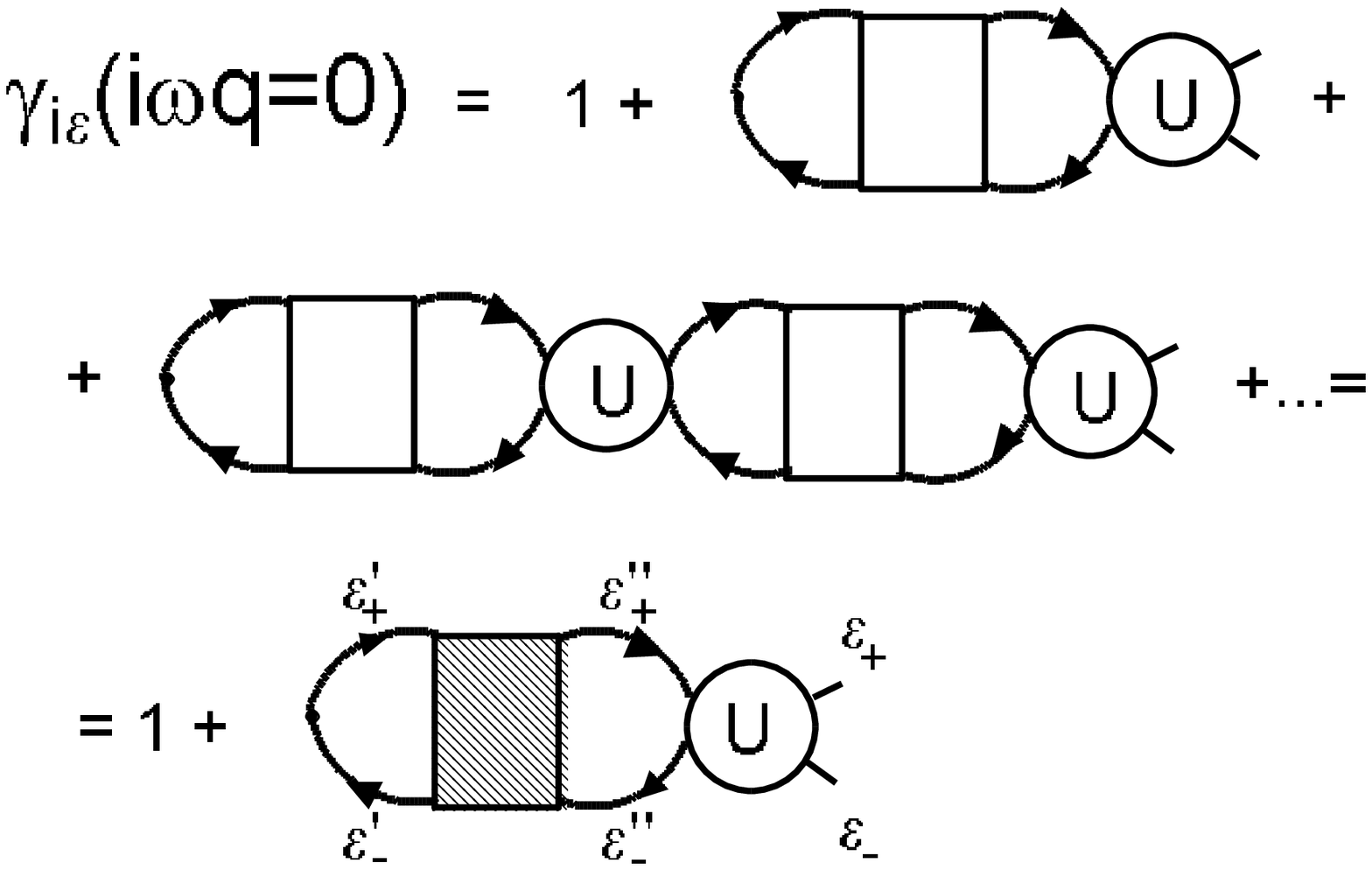}
\caption{Effective vertex $\gamma_{i\varepsilon}(i\omega,{\bf q}=0)$ used in 
calculations of conductivity.}
\label{gamma}
\end{figure}

As we noted in Ref. \cite{PRB07}, it is clear from (\ref{cond_gener})
that calculation of conductivity requires only the knowledge of 
$q^2$-contribution to $\Phi(i\omega,{\bf q})$. 
This can be easily found in the following way. 
First of all, note that all the loops in (\ref{BS_equ}) contain
$q$-dependence starting from terms of the order of $q^2$. Then we can take
an arbitrary loop (crossection) in the expansion of (\ref{BS_equ}) (see
Fig. \ref{BS_loop}), calculating it up to terms of the order of $q^2$, and 
make resummation of all contributions to the right and to the left from 
this crossection, putting $q=0$ in all these 
graphs. This is equivalent to simple $q^2$-differentiation of the expanded 
version of Eq. (\ref{BS_equ}).
This procedure immediately leads to the following relation for 
$q^2$-contribution to (\ref{PhiM}):
\begin{equation}
\phi(i\omega)\equiv\lim_{q\to 0}\frac{\Phi(i\omega,{\bf q})-
\Phi(i\omega,0)}{q^2}=\sum_{\varepsilon}\gamma^2_{i\varepsilon}(i\omega,{\bf q}=0)
\phi^0_{i\varepsilon}(i\omega)
\label{fi_func}
\end{equation}
where
\begin{equation}
\phi^0_{i\varepsilon}(i\omega)\equiv\lim_{q\to 0}
\frac{\Phi^0_{i\varepsilon}(i\omega,{\bf q})-
\Phi^0_{i\varepsilon}(i\omega,0)}{q^2}
\label{fi0_func}
\end{equation}
where $\Phi^0_{i\varepsilon}(i\omega,{\bf q})$
contains vertex corrections only due to impurity scattering,
while one-particle Green's functions entering it are taken
with self-energies due to both impurity scattering and local DMFT-like 
interaction, like in Eq. (\ref{Gk}). The vertex 
$\gamma_{i\varepsilon}(i\omega,{\bf q}=0)$ 
is determined diagrammatically as shown in Fig. \ref{gamma}, or analytically:
\begin{equation}
\gamma_{i\varepsilon}(i\omega,{\bf q}=0)=
1+\sum_{\varepsilon'\varepsilon''}
U_{i\varepsilon i\varepsilon''}(i\omega)
\Phi_{i\varepsilon''i\varepsilon'}(i\omega,{\bf q}=0).
\label{gamma_equ}
\end{equation}
Now using Bethe--Salpeter equation (\ref{BS_equ}) we can write explicitly:
\begin{eqnarray}
&&\gamma_{i\varepsilon}(i\omega,{\bf q}=0)=1+\nonumber\\
&&+\sum_{\varepsilon'}
\frac{\Phi_{i\varepsilon i\varepsilon'}(i\omega,{\bf q}=0)-
\Phi^0_{i\varepsilon}(i\omega,{\bf q}=0)}{\Phi^0_{i\varepsilon}(i\omega,{\bf q}=0)}=
\nonumber\\
&&=\frac{\sum_{\varepsilon'}\Phi_{i\varepsilon i\varepsilon'}(i\omega,{\bf q}=0)}
{\Phi^0_{i\varepsilon}(i\omega,{\bf q}=0)}.
\label{gamm_expl} 
\end{eqnarray}
For ${\bf q}=0$ we have the following Ward identity, which can be obtained by direct
generalization of the proof given in \cite{VW,Mig} 
(see details in Appendix of Ref. \cite{PRB07}):
\begin{eqnarray}
&&(-i\omega)\Phi_{i\varepsilon}(i\omega,{\bf q}=0)=
(-i\omega)\sum_{\varepsilon'}\Phi_{i\varepsilon i\varepsilon'}
(i\omega,{\bf q}=0)\nonumber\\
&&=\sum_{\bf p}G(i\varepsilon+i\omega,{\bf p})-
\sum_{\bf p}G(i\varepsilon,{\bf p}).
\label{Ward1}
\end{eqnarray}
Denominator of (\ref{gamm_expl}) contains vertex corrections only from
impurity scattering, while Green's functions
here are ``dressed'' {\em both} by impurities and local
(DMFT) Hubbard interaction. Thus we may consider the loop entering the
denominator as dressed by impurities only, but with ``bare''
Green's functions:
\begin{equation}
\tilde G_{0}(i\varepsilon,{\bf p})=\frac{1}{i\varepsilon+\mu
-\epsilon({\bf p})- \Sigma(i\varepsilon)},
\label{G_DMFT} 
\end{equation}
where $\Sigma(i\varepsilon)$ is local contribution to self-energy from DMFT.
For this problem we have the following Ward identity, similar to (\ref{Ward1}) 
(see Appendix of Ref. \cite{PRB07}):
\begin{eqnarray}
&&\sum_{\bf p}G(i\varepsilon+i\omega,{\bf p})-\sum_{\bf p}G(i\varepsilon,{\bf p})=
\nonumber\\
&&=\Phi_{i\varepsilon}^0(i\omega,{\bf q}=0)\left[\Sigma(i\varepsilon+i\omega)-
\Sigma(i\varepsilon)-i\omega\right]\equiv\nonumber\\
&&\equiv\Phi_{i\varepsilon}^0(i\omega,{\bf q}=0)\left[\Delta\Sigma(i\omega)-i\omega\right],
\label{Ward2}
\end{eqnarray}
where we have introduced 
\begin{equation}
\Delta\Sigma(i\omega)=\Sigma(i\varepsilon+i\omega)-
\Sigma(i\varepsilon).
\label{D_Sigma}
\end{equation}
Thus, using (\ref{Ward1}), (\ref{Ward2}) in (\ref{gamm_expl}) we get 
the final expression for $\gamma_{i\varepsilon}(i\omega,{\bf q}=0)$:  
\begin{equation} 
\gamma_{i\varepsilon}(i\omega,{\bf q}=0)=1-\frac{\Delta\Sigma(i\omega)}{i\omega}.
\label{gamm_fin}
\end{equation}
Then (\ref{fi_func}) reduces to:
\begin{equation}
\phi(i\omega)=\sum_{\varepsilon}\phi^0_{i\varepsilon}(i\omega)
\left[1-\frac{\Delta\Sigma(i\omega)}
{i\omega}\right]^2.
\label{fi_funct}
\end{equation}
Analytic continuation to real frequencies is obvious and using (\ref{fi_func}),
(\ref{fi_funct}) in (\ref{cond_gener}) we can write the final expression for 
the real part of dynamic (optical) conductivity as:
\begin{eqnarray}
&&{\rm{Re}}\sigma(\omega)=\frac{e^2\omega}{2\pi}
\int_{-\infty}^{\infty}d\varepsilon\left[f(\varepsilon_-)
-f(\varepsilon_+)\right]\nonumber\\
&&{\rm{Re}}\left\{\phi^{0RA}_{\varepsilon}(\omega)\left[1-
\frac{\Sigma^R(\varepsilon_+)-\Sigma^A(\varepsilon_-)}{\omega}\right]^2-
\right.\nonumber\\
&&\left.-\phi^{0RR}_{\varepsilon}(\omega)\left[1-
\frac{\Sigma^R(\varepsilon_+)-\Sigma^R(\varepsilon_-)}{\omega}\right]^2
\right\}.
\label{cond_final}
\end{eqnarray}
Thus we have achieved a great simplification of our problem. To calculate
dynamic conductivity in DMFT+$\Sigma$ approximation we only have to solve 
single--particle problem as described by DMFT+$\Sigma$ procedure
above to determine self--consistent values of local self--energies 
$\Sigma(\varepsilon_{\pm})$, while non-trivial contribution of impurity
scattering are to be included via (\ref{fi0_func}), which is to be 
calculated in some approximation, taking into account only interaction with
impurities (random scattering), but using the ``bare'' Green's
functions of the form (\ref{G_DMFT}), which include local self--energies
already determined via the general DMFT+$\Sigma$ procedure.
Actually (\ref{cond_final}) provides also an effective algorithm to
calculate dynamic conductivity in standard DMFT (neglecting impurity
scattering), as (\ref{fi0_func}) is then easily calculated from a simple
loop diagram, determined by two Green's functions and free {\em scalar} 
vertices.  As usual, there is no need to calculate vertex corrections within 
DMFT itself, as was proven first considering the loop with {\em vector} 
vertices \cite{pruschke,georges96}. Obviously, Eq. (\ref{cond_final}) provides
effective interpolation between the case of strong correlations without disorder
and the case of pure disorder, without Hubbard correlations, which is of
major interest to us. In the following we shall see that calculations based
on Eq. (\ref{cond_final}) give a reasonable overall picture of MIT in
Anderson--Hubbard model.

\subsection{Self-consistent equations for generalized diffusion
coefficient and conductivity}

\begin{figure}
\includegraphics[clip=true,width=0.35\textwidth]{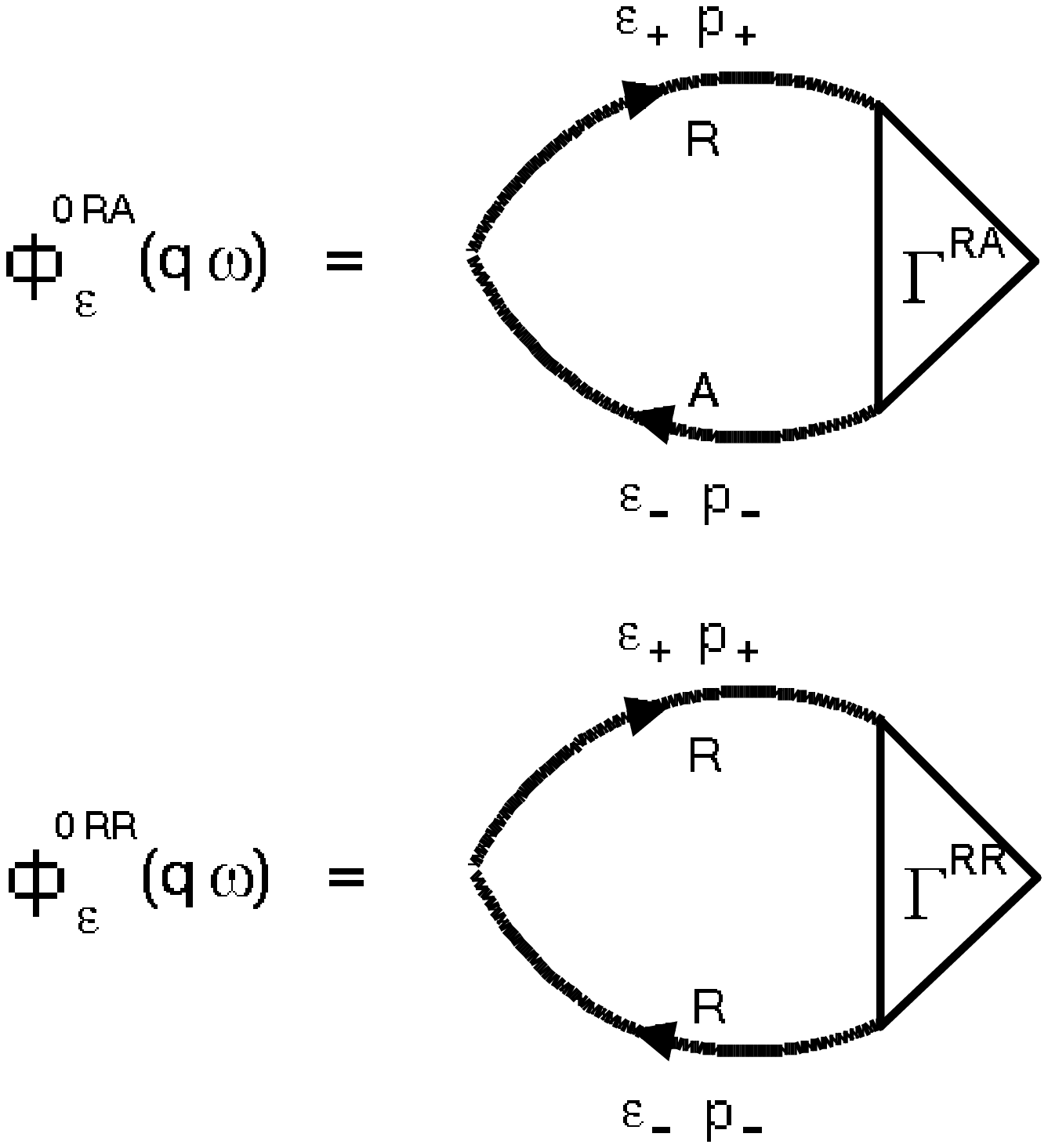}
\caption{Diagrammatic representation of 
$\Phi^{0RA}_{\varepsilon}(\omega,{\bf q})$ and $\Phi^{0RR}_{\varepsilon}
(\omega,{\bf q})$.} 
\label{loop_fi} 
\end{figure}

Now to calculate optical conductivity we need the knowledge of the 
basic block $\Phi^0_{i\varepsilon}(i\omega,{\bf q})$, entering 
(\ref{fi0_func}), or, more precisely, appropriate functions analytically 
continued to real frequencies: $\Phi^{0RA}_{\varepsilon}(\omega,{\bf q})$ and
$\Phi^{0RR}_{\varepsilon}(\omega,{\bf q})$, which in turn define
$\phi^{0RA}_{\varepsilon}(\omega)$ and $\phi^{0RR}_{\varepsilon}(\omega)$
entering (\ref{cond_final}), and defined by obvious relations similar to
(\ref{fi0_func}): 
\begin{equation}
\phi^{0RA}_{\varepsilon}(\omega)=\lim_{q\to 0}
\frac{\Phi^{0RA}_{\varepsilon}(\omega,{\bf q})-
\Phi^{0RA}_{\varepsilon}(\omega,0)}{q^2},
\label{fi0RA_func}
\end{equation}
\begin{equation}
\phi^{0RR}_{\varepsilon}(\omega)=\lim_{q\to 0}
\frac{\Phi^{0RR}_{\varepsilon}(\omega,{\bf q})-
\Phi^{0RR}_{\varepsilon}(\omega,0)}{q^2}
\label{fi0RR_func}.
\end{equation}
By definition we have:
\begin{eqnarray}
\Phi^{0RA}_{\varepsilon}(\omega,{\bf q})=\sum_{\bf p}
G^R(\varepsilon_+,{\bf p_+})G^A(\varepsilon_-,{\bf p_-})\times\nonumber\\
\times\Gamma^{RA}(\varepsilon_-,{\bf p}_-;\varepsilon_+,{\bf p}_+)
\label{PhiRA}
\end{eqnarray}
\begin{eqnarray}
\Phi^{0RR}_{\varepsilon}(\omega,{\bf q})=\sum_{\bf p}
G^R(\varepsilon_+,{\bf p_+})G^R(\varepsilon_-,{\bf p_-})\times\nonumber\\
\times\Gamma^{RR}(\varepsilon_-,{\bf p}_-;\varepsilon_+,{\bf p}_+),\\ \nonumber
\label{PhiRR}
\end{eqnarray}
which are shown diagrammatically in Fig. \ref{loop_fi} 
and ${\bf p}_{\pm}={\bf p}\pm\frac{\bf q}{2}$. 
Here Green's functions $G^R(\varepsilon_+,{\bf p_+})$ and
$G^A(\varepsilon_-,{\bf p_-})$ are defined by analytic continuation 
$(i\varepsilon\to\varepsilon\pm i\delta)$ of Matsubara Green's functions 
(\ref{Gk}) determined via our DMFT+$\Sigma$ algorithm (\ref{Gloc}) -- 
(\ref{BornSigma}), while vertices 
$\Gamma^{RA}(\varepsilon_-,{\bf p}_-;\varepsilon_+,{\bf p}_+)$ and
$\Gamma^{RR}(\varepsilon_-,{\bf p}_-;\varepsilon_+,{\bf p}_+)$ contain all
vertex corrections due to impurity scatterings.

The most important block $\Phi^{0RA}_{\varepsilon}(\omega,{\bf q})$ can be 
calculated using the basic approach of self--consistent theory of localization 
\cite{VW,WV,MS,MS86,VW92,Diagr} with appropriate extensions, taking into 
account the role of the local Hubbard interaction using DMFT+$\Sigma$ 
approach. The only important difference with the standard approach
is that equations of self--consistent theory are now derived using  
\begin{equation}
G^{R,A}(\varepsilon,{\bf p})=\frac{1}{\varepsilon+\mu-\epsilon({\bf p})-
\Sigma^{R,A}(\varepsilon)
-\Sigma^{R,A}_{imp}(\varepsilon)}
\label{GkE}
\end{equation}
containing DMFT contributions $\Sigma^{R,A}(\varepsilon)$, not only impurity
scattering contained in:
\begin{equation}
\Sigma_{imp}^{R,A}(\varepsilon)=\Delta^2\sum_{\bf p}G^{R,A}(\varepsilon,{\bf p}) 
={\rm Re}\Sigma_{imp}(\varepsilon)\pm i\gamma(\varepsilon) 
\label{BornSigmaE} 
\end{equation}
where $\gamma(\varepsilon)=\pi\Delta^2N(\varepsilon)$ 
and $N(\varepsilon)$ is the density of states renormalized by Hubbard 
interaction, accounted via DMFT+$\Sigma$ and given as usual by:
\begin{equation}
N(\varepsilon)=-\frac{1}{\pi}\sum_{\bf p}{\rm Im}G^{R}(\varepsilon,{\bf p})
\label{DSt}
\end{equation} 

Following all the usual steps of standard derivation 
\cite{VW,WV,MS,MS86,VW92,Diagr} we obtain diffusion like 
(at small $\omega$ and $q$) contribution to
$\Phi^{0RA}_{\varepsilon}(\omega,{\bf q})$ as:
\begin{equation}
\Phi^{0RA}_{\varepsilon}({\bf q},\tilde\omega)=\frac{2\pi iN(\varepsilon)}
{\tilde\omega+iD(\omega)q^2} 
\label{FiRA} 
\end{equation}
where important difference with the single--particle case is contained in
\begin{eqnarray}
\tilde\omega=\varepsilon_+-\varepsilon_-
-\Sigma^R(\varepsilon_+)+\Sigma^A(\varepsilon_-)=\nonumber\\
=\omega-\Sigma^R(\varepsilon_+)+\Sigma^A(\varepsilon_-)\equiv
\omega-\Delta\Sigma^{RA}(\omega)
\label{tomega}
\end{eqnarray}
which replaces the usual $\omega$ term in the denominator of the standard
expression for $\Phi^{0RA}_{\varepsilon}(\omega,{\bf q})$. On general grounds
it is clear that in metallic phase for $\omega\to 0$ we have 
$\Delta\Sigma^{RA}(\omega=0)=2{\rm iIm}\Sigma(\varepsilon)\sim Max\{T^2,\varepsilon^2\}$,
reflecting Fermi--liquid behavior of DMFT (conserved by elastic impurity
scattering). At finite $T$ it leads to the usual phase decoherence due to
electron -- electron scattering \cite{Lee85,ma}.
Generalized diffusion coefficient $D(\omega)$ will be determined by
solving the basic self--consistency equation, introduced below.

Now using (\ref{FiRA}) in (\ref{fi0RA_func}) we easily obtain:
\begin{equation}
\phi^{0RA}_{\varepsilon}(\omega)=
\frac{2\pi N(\varepsilon)D(\omega)}
{\omega^2\left(1-\frac{\Delta\Sigma^{RA}(\omega)}{\omega}\right)^2}
\label{fi0_fnc}
\end{equation}
Then using (\ref{fi0_fnc}) in (\ref{cond_final}), for $\omega\to 0$ and 
for $T=0$ we get just the usual Einstein relation for static conductivity:
\begin{equation}
\sigma(0)=e^2N(0)D(0)
\label{st_cond}
\end{equation}
All contributions form Hubbard interaction are reduced to renormalization of
the density of states at the Fermi level and also of diffusion coefficient
$D(0)$.

Then (\ref{cond_final}) reduces to:
\begin{eqnarray}
&& {\rm{Re}}\sigma(\omega)=\frac{e^2\omega}{2\pi}
\int_{-\infty}^{\infty}d\varepsilon\left[f(\varepsilon_-)
-f(\varepsilon_+)\right]\times\nonumber\\
&&{\rm Re}\left\{\frac{2\pi N(\varepsilon)D(\omega)}
{\omega^2}-\phi^{0RR}_{\varepsilon}(\omega)\left[1-
\frac{\Delta\Sigma^{RR}(\omega)}{\omega}\right]^2\right\},
\nonumber\\
\label{con_fin}
\end{eqnarray}
where the second term actually can be neglected at small $\omega$, or
just calculated from (\ref{fi0RR_func}) taking 
$\Phi^{0RR}_{\varepsilon}(\omega,{\bf q})$  given by the usual ``ladder'' 
approximation (\ref{Ladder}).   

Now we have to formulate our basic self--consistent equation, determining
the generalized diffusion coefficient $D(\omega)$. Again we follow all the 
usual steps of self--consistent theory of localization (see details in 
the Appendix A), taking into account the form of our single--particle Green's 
function (\ref{GkE}), and not restricting analysis to small $\omega$ limit. 
Then we can write the generalized diffusion coefficient as:  
\begin{equation} 
D(\omega)=\frac{<v>^2}{d}\frac{i}{\tilde\omega+M(\omega)}
\label{Dgen}
\end{equation}
where $d$ is spatial dimensionality and average velocity $<v>$ is defined in
(\ref{veloc1}) (to a good approximation it is just the Fermi velocity),
while the relaxation kernel $M(\omega)$ satisfies self--consistency
equation, similar to that derived in Refs. \cite{VW,WV,MS,MS86,VW92,Diagr},
using ``maximally crossed'' diagrams for irreducible impurity scattering vertex
(built with Green's functions (\ref{GkE})):
\begin{equation}
M(\omega)=-\Delta\Sigma_{imp}^{RA}(\omega)
+\Delta^4\sum_{\bf p}(\Delta G_{\bf p})^2\sum_{\bf q}
\frac{1}{\tilde\omega+iD(\omega)q^2}
\label{MKernel}
\end{equation}
with
\begin{equation}
\Delta G_{\bf p}=G^R(\varepsilon_+,{\bf p})-G^A(\varepsilon_-,{\bf p})
\label{DGp}
\end{equation}
and $\Delta\Sigma^{RA}_{imp}(\omega)=\Sigma_{imp}^R(\varepsilon_+)
-\Sigma_{imp}^A(\varepsilon_-)$ is due to impurity scattering.
It is important to stress once again that there are no contributions to this
equation due to vertex corrections, determined by local Hubbard interaction.
Using the definition (\ref{Dgen}) Eq. (\ref{MKernel}) can be rewritten as 
self--consistent equation for the generalized diffusion coefficient itself:
\begin{eqnarray}
D(\omega)=i\frac{<v>^2}{d}\Bigg\{\tilde\omega-
\Delta\Sigma_{imp}^{RA}(\omega)+\nonumber\\
\left.+\Delta^4\sum_{\bf p}(\Delta G_{\bf p})^2\sum_{\bf q}
\frac{1}{\tilde\omega+iD(\omega)q^2}\right\}^{-1}
\label{Dsc}
\end{eqnarray}
which is to be solved in conjunction with our DMFT+$\Sigma$ loop 
(\ref{Gk})--(\ref{BornSigma}).
Due to the limits of diffusion approximation, summation over $q$ in (\ref{Dsc})
is to be restricted to:
\begin{equation}
q<k_0=Min \{l^{-1},p_F\}
\label{cutoff}
\end{equation}
where $l=<v>/2\gamma(0)$ is an elastic mean--free path, $p_F$ is the 
Fermi momentum \cite{MS86,Diagr}. 

Solving (\ref{Dsc}) for different sets of parameters of our model and 
using it in (\ref{con_fin}) with regular contributions from (\ref{Ladder}) 
we can calculate dynamic (optical) conductivity in different phases of 
Anderson--Hubbard model.

\section{Results and discussion}
\label{results}

We performed extensive numerical calculations for simplified version of 
three--dimensional Anderson--Hubbard model on cubic lattice with semi--elliptic 
DOS of ``bare'' band of the width $W=2D$:
\begin{equation}
N_0(\varepsilon)=\frac{2}{\pi D^2}\sqrt{D^2-\varepsilon^2}
\label{DOS}
\end{equation}
DOS is always given in units of number of states per energy interval,
per lattice cell volume $a^3$ ($a$ is lattice spacing), per spin. Some 
related technical details are given in Appendix B.

Mostly we shall concentrate on the half -- filled case, though some results for
finite dopings will be also presented. Fermi level is always placed at zero
energy. 

As ``impurity solver'' of DMFT we employed the
reliable numerically exact method of numerical renormalization group (NRG) 
\cite{NRG,BPH,Bull}.  Calculations were done for temperatures $T\sim 0.001D$,
which effectively makes temperature effects in DOS and conductivity
negligible. Discretization parameter of NRG was always
$\Lambda$=2, number of low energy states after truncation 1000, cut-off near
Fermi energy $!)^{-6}$, broadening parameter $b$=0.6.   

Below we present only a fraction of most typical results.

\subsection{Density of states evolution}

Within the standard DMFT approach density of states of the half--filled
Hubbard model has a typical three peak structure: a narrow quasiparticle
band (central peak)
develops at the Fermi level, with wider upper and lower Hubbard bands 
forming at $\varepsilon\sim\pm U/2$. Quasiparticle band narrows further  
with the growth of $U$ in metallic phase, vanishing at critical $U_{c2}\approx
1.5W$, signifying the Mott--Hubbard MIT with a gap opening at the Fermi level
\cite{pruschke,georges96,Bull}.

\begin{figure}
\includegraphics[clip=true,width=0.5\textwidth]{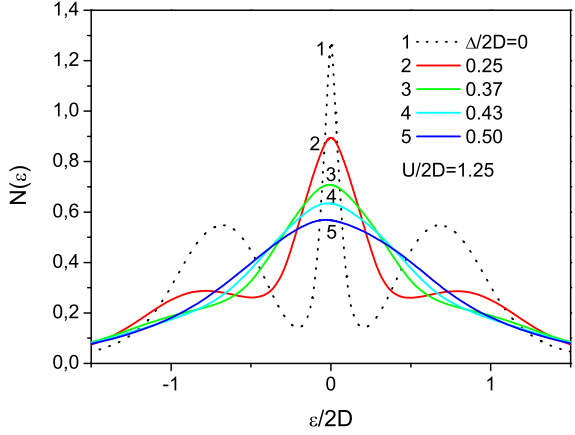}
\caption{(Color online) Density of states of half--filled Anderson--Hubbard 
model for different degrees of disorder $\Delta$, and $U=2.5D$, typical for
correlated metal.} 
\label{metDOS} 
\end{figure}

In Fig. \ref{metDOS} we present our DMFT+$\Sigma$ results for the
density of states, obtained for $U=2.5D=1.25W$ typical for correlated metal
without disorder, for different degrees of disorder $\Delta$, including strong 
enough values, actually transforming correlated metal to correlated Anderson 
insulator (see next subsection \ref{cond}). 
As may be expected, we observe typical widening and damping of DOS by disorder.

\begin{figure}
\includegraphics[clip=true,width=0.5\textwidth]{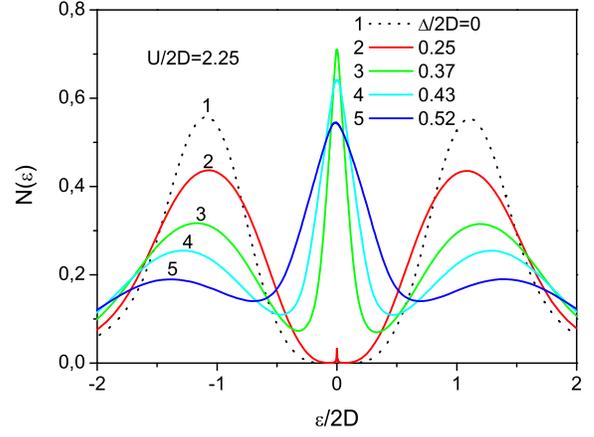}
\caption{(Color online) Density of states of half--filled Anderson--Hubbard model for
different degrees of disorder $\Delta$ and $U=4.5D$, 
typical for Mott insulator.} 
\label{insDOS} 
\end{figure}

More unexpected are the results obtained for the values of $U$ typical for
Mott insulator without disorder, as shown in Fig. \ref{insDOS} for $U=4.5D
=2.25W$.
We see the restoration of central peak (quasiparticle band) in DOS with the
growth of disorder, transforming Mott insulator either to correlated metal or
correlated Anderson insulator. Similar behavior of DOS was recently obtained 
in Ref. \cite{BV}. However, in our calculations the presence of distinct 
Hubbard bands was observed even for rather large values of disorder, with no 
signs of vanishing Hubbard structure of DOS,  which was observed in  
of Ref. \cite{BV}. This is probably due to very simple nature of our
approximation for DOS under disordering, though we must stress that this
difference may be also due to another model of disorder used in Ref. \cite{BV}, 
i.e. flat distribution of $\epsilon_i$ in (\ref{And_Hubb}) instead of our 
Gaussian case (\ref{Gauss}). Though unimportant, in general, to physics of 
Anderson transition, the type of disorder may be significant for the DOS 
behavior.

It is well known, that hysteresis behavior of DOS is obtained for 
Mott--Hubbard transition if we perform DMFT calculations with $U$ decreasing
from insulating phase \cite{georges96,Bull}. Mott insulator phase survives 
for the values of $U$ well inside the correlated metal phase, obtained with
the increase of $U$. Metallic phase is restored at $U_{c1}\approx 1.0W$. The
values of $U$ from the interval $U_{c1}<U<U_{c2}$ are usually considered 
as belonging to coexistence region of metallic and (Mott) insulating phases,
with metallic phase being thermodynamically more stable 
\cite{georges96,Bull,Blum}.

\begin{figure}
\includegraphics[clip=true,width=0.5\textwidth]{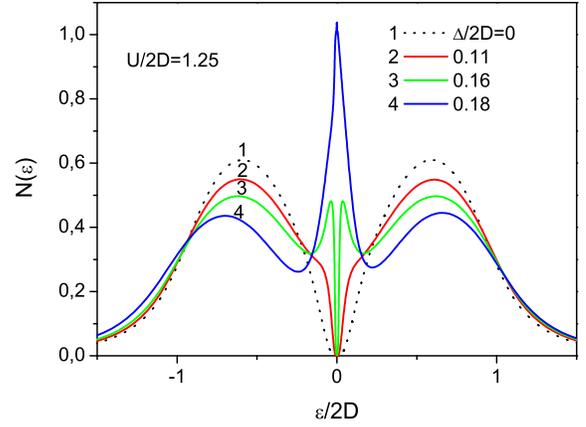}
\caption{(Color online) Restoration of quasiparticle band by disorder in 
coexistence (hysteresis) region for $U=2.5D$, obtained from Mott insulator with 
decreasing $U$.} 
\label{hystDOS} 
\end{figure}

In Fig. \ref{hystDOS} we present our typical data for DOS with different
disorder for the same value of $U=2.5D=1.25W$ as in Fig. \ref{metDOS}, but 
for hysteresis region, obtained by decreasing $U$
from Mott insulator phase. Again we observe the restoration of central peak
(quasiparticle band) in DOS under disordering. Note also the peculiar form
of DOS around the Fermi level during this transition -- a narrow energy gap
is conserved until it is closed by disorder, and central
peak is formed from two symmetric maxima in DOS joining into quasiparticle
band. This reminds similar behavior observed in periodic Anderson model
\cite{georges96}. Apparently this effect was unnoticed in previous 
calculations of DOS in coexistence region \cite{Bull} (in the absence of 
disorder), while in our case it was obtained mainly due to our use of very 
fine mesh of the values of disorder parameter $\Delta$.

The physical reason for rather unexpected restoration of the central 
(quasiparticle) peak in DOS is in fact clear. The controlling parameter
for its appearance or disappearance in DMFT is actually the ratio of Hubbard
interaction $U$ and the bare bandwidth $W=2D$. Under disordering we obtain the
new effective bandwidth $W_{eff}$ (in the absence of Hubbard interaction) 
which grows with disorder, while semi--elliptic form of the
DOS, with well--defined band edges is conserved in self--consistent 
Born approximation (\ref{BornSigma}). This leads to the decrease of the ratio
$U/W_{eff}$, which induces the reappearance of quasiparticle band in our model.
This will be illustrated in more detail in subsection \ref{phd}, where
our DOS calculations within DMFT+$\Sigma$ approach for a wide range
of parameters will be used to study the phase diagram of Anderson--Hubbard
model.

\subsection{Dynamic conductivity: Mott--Hubbard and Anderson transitions}
\label{cond}

Real part of dynamic (optical) conductivity was calculated for different 
combinations of parameters of our model directly from Eqs. (\ref{con_fin}), 
(\ref{Ladder}) and (\ref{Dsc}) using the results of DMFT+$\Sigma$ loop 
(\ref{Gk}) -- (\ref{BornSigma}) as an input. The values of conductivity below
are given in natural units of ${e^2}/{\hbar a}$ ($a$ -- lattice spacing).

In the absence of disorder we obviously reproduce the results of the standard
DMFT \cite{pruschke,georges96} with dynamic conductivity characterized in
general by the usual (metallic) Drude--like peak at zero frequency and wide 
absorption maximum at $\omega\sim U$, corresponding to transitions to the 
upper Hubbard band. With the growth of $U$ Drude peak decreases and vanishes
at Mott transition, when only transitions through the Mott--Hubbard gap
contribute.
Introduction of disorder leads to qualitative changes in the frequency
dependence of conductivity. Below we mainly show the results obtained
for the same values of $U$ and $\Delta$ as were used above to illustrate
DOS behavior. 

\begin{figure}
\includegraphics[clip=true,width=0.5\textwidth]{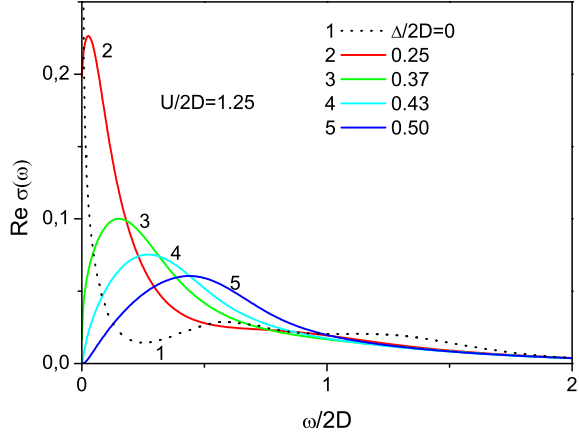}
\caption{(Color online) Real part of dynamic conductivity for half--filled 
Anderson--Hubbard model for different degrees of disorder $\Delta$, 
and $U=2.5D$, typical for correlated metal. Lines 1,2 are for metallic
phase, line 3 corresponds to the mobility edge (Anderson transition), 
lines 4,5 correspond to correlated Anderson insulator.} 
\label{met_cond} 
\end{figure}

In Fig. \ref{met_cond} we present real part of dynamic (optical) 
conductivity for half--filled Anderson--Hubbard model for different degrees 
of disorder $\Delta$, and $U=2.5D$, typical for correlated metal. Transitions
to upper Hubbard band at $\omega\sim U$ are practically unobservable in these 
data. However, it is clearly seen that metallic Drude peak at zero frequency 
is widened and suppressed, being gradually transformed to a peak at finite 
frequncies due to effects of 
Anderson localization. Anderson transition takes place at $\Delta_c\approx 
0.74D=0.37W$ (which in all our graphs (including those for DOS) corresponds to 
curve 3). Note that this value is actually dependent on the value of
cutoff (\ref{cutoff}), which is defined up to a constant of the order of
unity \cite{MS86,Diagr}. Naive expectations may have lead to a
conclusion, that a narrow quasiparticle band at the Fermi level, which forms in 
general case of highly correlated metal, may be localized much more easily than
typical conduction band. However, we see that these expectations are just
wrong and this band is localized only at strong enough disorder 
$\Delta_c\sim D$, the same as for the whole conduction band of the width 
$\sim W$. This is in accordance with previous analysis of localization in
two-band model \cite{ErkS}.

More important is the fact that in
DMFT+$\Sigma$ approximation the value of $\Delta_c$ is independent of $U$
as all interaction effects enter Eq. (\ref{Dsc}) only via 
$\Delta\Sigma^{RA}(\omega)\to 0$ for $\omega\to 0$ (at $T=0$), so that 
interaction just drops out at $\omega=0$. This is actually the main deficiency 
of our approximation, which is due to our neglect of interference effects between 
interaction and disorder scattering. Important role of these interference
effects is known for a long time \cite{Lee85,ma}. However, despite the neglect
of these effects we are able to produce physically sound interpolation between 
two main limits of interest -- pure Anderson transition due to disorder and 
Mott--Hubbard transition due to strong correlations. Thus we consider it as a 
reasonable first step to future complete theory of MIT in strongly correlated 
disordered systems.  
 
\begin{figure}
\includegraphics[clip=true,width=0.5\textwidth]{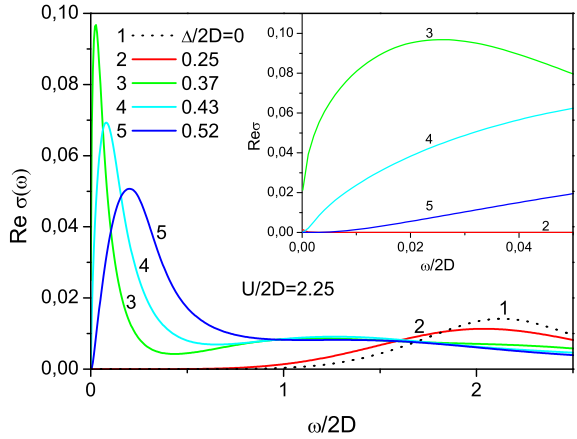}
\caption{(Color online) Real part of dynamic conductivity of half--filled 
Anderson--Hubbard model for different degrees of disorder $\Delta$ and $U=4.5D$, 
typical for Mott insulator. Lines 1,2 correspond to Mott insulator,
line 3 corresponds to the mobility edge (Anderson transition), lines 4,5 are 
for correlated Anderson insulator. Insert -- enlarged region of small 
frequencies.} 
\label{ins_cond} 
\end{figure}

In Fig. \ref{ins_cond} we present real part of dynamic (optical) 
conductivity for different degrees 
of disorder $\Delta$, and $U=4.5D$, typical for Mott--Hubbard insulator. 
At the insert we show our data for small frequencies, which allow clear
distinction of different types of conductivity behavior, especially close
to Anderson transition, or in Mott insulator phase.
In this figure we clearly see the contribution of transitions to upper Hubbard 
band at $\omega\sim U$. More importantly we observe,
that the growth of disorder produces finite conductivity
within the frequency range of Mott--Hubbard gap, which correlates with the
appearance of quasiparticle band (central peak) in DOS within this gap,
as shown in Fig. \ref{insDOS}. In general case, this conductivity is metallic 
(finite in the static limit of $\omega=0$) for $\Delta<\Delta_c$, while for
$\Delta>\Delta_c$ at small frequencies we obtain 
${\rm Re}\sigma(\omega)\sim\omega^2$, which is
typical of Anderson insulator \cite{VW,WV,MS,MS86,VW92,Diagr}. Note that due
to a finite internal accuracy of NRG numerics, small but finite spurious
contributions to ${\rm Im}\Sigma^{R,A}(\varepsilon=0)$ always appear \cite{Bull}
and formally grow with $U$. These contributions are practically irrelevant in 
calculations of conductivity in metallic state. However, in Anderson 
insulator these spurious terms contribute via $\tilde\omega$ in Eq. 
(\ref{Dsc}) and lead to unphysical finite dephasing effects at $\omega=0$ (or 
$T=0$), which can simulate small finite static conductivity. To exclude these
spurious effects we had to make appropriate subtractions in our data for
${\rm Im}\Sigma^{R,A}(\varepsilon)$ at $\varepsilon=0$.

\begin{figure}
\includegraphics[clip=true,width=0.5\textwidth]{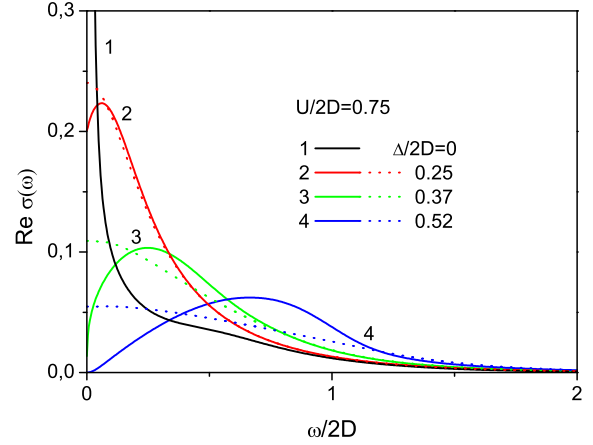}
\caption{(Color online) Real part of dynamic conductivity of half--filled 
Anderson--Hubbard model for different degrees of disorder $\Delta$ and $U=1.5D$, 
comparison of self--consistent theory (full lines) with ``ladder'' 
approximation (dotted lines).} 
\label{comp_ladd} 
\end{figure}

Rather unusual is the appearance of low frequency peak in 
${\rm Re}\sigma(\omega)$ at low frequencies even in metallic phase.
This is due to importance of weak localization effects, as can be clearly
seen from Fig. {\ref{comp_ladd}}, where we compare the real part of dynamic 
conductivity for different degrees of disorder $\Delta$ and $U=1.5D$, 
obtained via our self--consistent approach (taking into account localization
effects via ``maximally crossed'' diagrams) with that obtained using the
``ladder'' approximation for $\Phi^{0RA}_{\varepsilon}(\omega,{\bf q})$
(similar to (\ref{Ladder})), which neglects all localization effects. It is
clearly seen that in this simple approximation we just obtain the usual
Drude--like peak at $\omega=0$, while the account of localization effects
produce the peak in ${\rm Re}\sigma(\omega)$ at low (finite) frequencies. 
Metallic state is defined \cite{Mott90} by the finite value of zero temperature
conductivity at $\omega=0$.

\begin{figure}
\includegraphics[clip=true,width=0.5\textwidth]{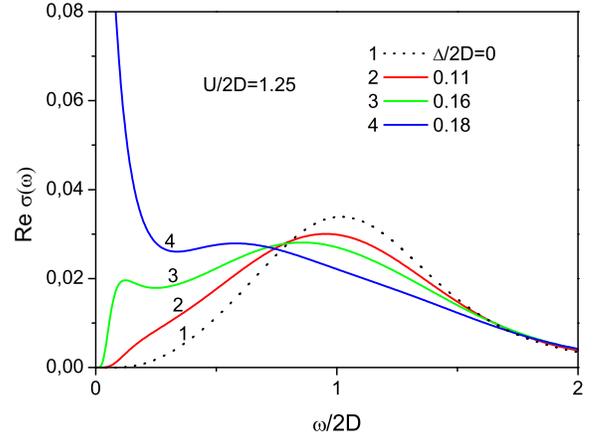}
\caption{(Color online) Real part of dynamic conductivity of half--filled 
Anderson--Hubbard model for different degrees of disorder $\Delta$ and $U=2.5D$, 
obtained from Mott insulator with decreasing $U$.} 
\label{con_coex} 
\end{figure}

Up to now we presented only conductivity data obtained with increase of $U$
from metallic to (Mott) insulating phase. 
As we decrease $U$ from Mott insulator
hysteresis of conductivity is observed in coexistence region, defined
(in the absence of disorder, $\Delta=0$) by $U_{c1}<U<U_{c2}$. Typical data
are shown in Fig. \ref{con_coex}, where we present the real part of dynamic 
conductivity for different degrees of disorder $\Delta$ and $U=2.5D$, 
obtained from Mott insulator with decreasing $U$, which should be compared
with those shown in Fig. \ref{met_cond}. Transition to metallic state via 
the closure of a narrow gap, ``inside'' much wider Mott--Hubbard gap, is 
clearly seen, which correlates with DOS data shown in Fig. \ref{hystDOS}.

\subsection{Phase diagram of half--filled Anderson--Hubbard model.}
\label{phd}

Phase diagram of half--filled Anderson--Hubbard model was studied in 
Ref. \cite{BV}, using the approach, based on direct DMFT calculations for a 
set of random realizations of site energies $\epsilon_i$ in (\ref{And_Hubb}) 
with subsequent averaging to get both the standard average DOS and also
geometrically averaged local DOS, which was used to determine transition to
Anderson insulator phase. Here we present our results for the 
zero--temperature phase diagram of half--filled paramagnetic 
Anderson--Hubbard model, obtained from extensive calculations of both average 
DOS and dynamic (optical) conductivity in DMFT+$\Sigma$ approximation.  
It should be noted, that conductivity calculations are the most direct way 
to distinguish metallic and insulating phases \cite{Mott90}.

\begin{figure}
\includegraphics[clip=true,width=0.5\textwidth]{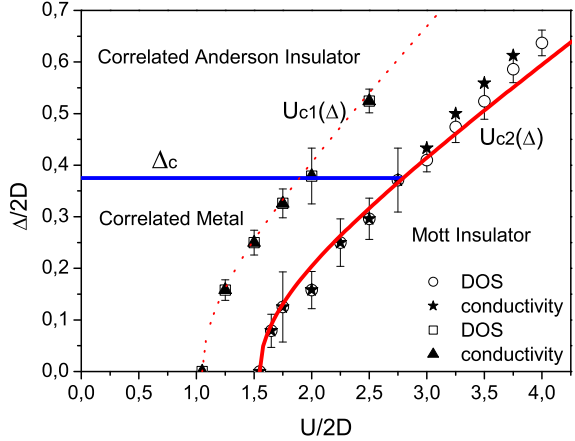}
\caption{(Color online) Zero temperature phase diagram of paramagnetic 
Anderson--Hubbard model. Borders of Mott insulator phase $U_{c1,c2}(\Delta)$
are shown as obtained from Eqs. (\ref{Uc}), while different symbols show 
values calculated either from DOS or conductivity behavior. 
Dotted line defines the border of coexistence region obtained with decreasing 
$U$ from Mott insulator phase. Line of Anderson transition is given by 
calculated value of $\Delta_c=0.37$.} 
\label{ph_diag} 
\end{figure}

Our phase diagram in the disorder--correlation $(\Delta,U)$--plane  
is shown in Fig. \ref{ph_diag}.  Anderson transition line $\Delta_c\approx
0.37W=0.74D$ was determined as the value of disorder for which the static 
conductivity becomes zero at $T=0$.  Mott--Hubbard transition may be 
determined either via the disappearance of central peak (quasiparticle band)
in DOS, or from conductivity, e.g. looking for the closure of gap in
dynamic conductivity in insulating phase, or from vanishing static conductivity
in metallic region. All these methods were used and appropriate results are 
shown for comparison in Fig. \ref{ph_diag}.

We already stressed that DMFT+$\Sigma$ approximation gives the universal
($U$--independent) value of $\Delta_c$. This is due to neglect of 
interference between disorder scattering and Hubbard interaction, and leads
to the main (over) simplification of our phase diagram, compared with that
obtained in Ref.\cite{BV}.  
Note, that direct comparison of our critical disorder
value with those of Ref. \cite{BV} is complicated by different types of 
random site--energies distributions used here (Gaussian) and in Ref. \cite{BV}
(rectangular distribution). As a rule of thumb (cf. second reference in
\cite{MS}) our Gaussian value of $\Delta_c$ should be multiplied by 
$\sqrt{12}$  to obtain the critical disorder value for rectangular 
distribution. This gives $\Delta_c\approx 1.28$ in rather good agreement with 
$\Delta_c(U=0)\approx 1.35W$ value of Ref. \cite{BV}, justifying our cutoff 
choice in (\ref{cutoff}).

The influence of disorder scattering on
Mott--Hubbard transition is highly non--trivial and in some respects is in 
qualitative agreement with results of Ref. \cite{BV}. 
The main difference is that our
data indicate the survival of Hubbard bands structures in DOS even in the
limit of rather large disorder, while these were claimed to disappear in
Ref. \cite{BV}. Also we obtain the coexistence region smoothly widening with 
the growth of disorder and not disappearing at some ``critical'' point, as in 
Ref. \cite{BV}. The borders of our coexistence region, which in fact define the
borders of Mott insulator phase obtained with increasing or decreasing $U$, 
are determined by the lines of $U_{c1}(\Delta)$ and $U_{c2}(\Delta)$ shown in 
Fig. \ref{ph_diag}, which are obtained from the simple equation:
\begin{equation}
\frac{U_{c1,c2}(\Delta)}{W_{eff}}=\frac{U_{c1,c2}}{W}
\label{UcW}
\end{equation} 
with
\begin{equation}
W_{eff}=W\sqrt{1+16\frac{\Delta^2}{W^2}}
\label{Weff}
\end{equation}
which is the effective bandwidth in the presence of disorder, calculated for
$U=0$ in self--consistent Born approximation (\ref{BornSigma}). Thus 
the borders of coexistence region are given by:
\begin{equation}
U_{c1,c2}(\Delta)=U_{c1,c2}\sqrt{1+16\frac{\Delta^2}{W^2}}
\label{Uc}
\end{equation}
which are explicitly shown in Fig. \ref{ph_diag} by dotted and full lines,
defining the borders of Mott insulator phase. Numerical
results for disappearance of quasiparticle band (central peak) in DOS, 
as well as points following from qualitative change in 
conductivity behavior, are shown in Fig. \ref{ph_diag} by different symbols 
demonstrating very good agreement with these 
lines, confirming the ratio (\ref{UcW}) as controlling parameter of Mott 
transition in the presence of disorder.

\begin{figure}
\includegraphics[clip=true,width=0.5\textwidth]{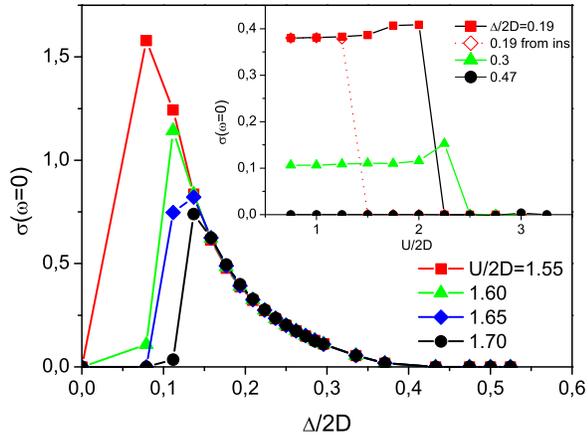}
\caption{(Color online) Disorder dependence of static conductivity, 
obtained for several values of $U$ and showing disorder induced Mott 
insulator to metal transition. At the insert we show static conductivity 
dependence on $U$ close to Mott transition, including typical
hysteresis behavior obtained with $U$ decreasing from Mott insulator phase.} 
\label{ins_met_dis} 
\end{figure}

Most striking result of our analysis (also qualitatively demonstrated in 
Ref. \cite{BV}) is the possibility of metallic state being restored from
Mott--Hubbard insulator with the growth of disorder. This is clear from the
phase diagram and is nicely demonstrated by our data for (static) conductivity 
shown in Fig. \ref{ins_met_dis} for several values of $U>U_{c2}$ and disorder 
values $\Delta<\Delta_c$. At the insert in Fig. \ref{ins_met_dis} we also
illustrate static conductivity hysteresis, observed in coexistence region of 
the phase diagram, obtained with $U$ decreasing from Mott insulator phase.

\subsection{Doped Mott insulator}

\begin{figure}
\includegraphics[clip=true,width=0.5\textwidth]{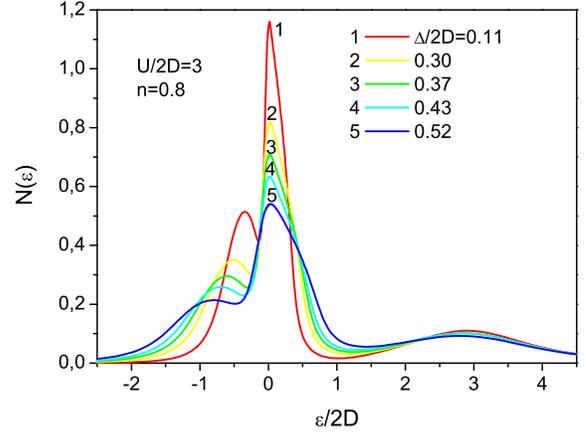}
\caption{(Color online) Density of states of Anderson--Hubbard model with 
electron concentration $n=0.8$ for different degrees of disorder $\Delta$ and 
$U=6.0D$, representing the doped Mott insulator.} 
\label{dopDOS} 
\end{figure}

\begin{figure}
\includegraphics[clip=true,width=0.5\textwidth]{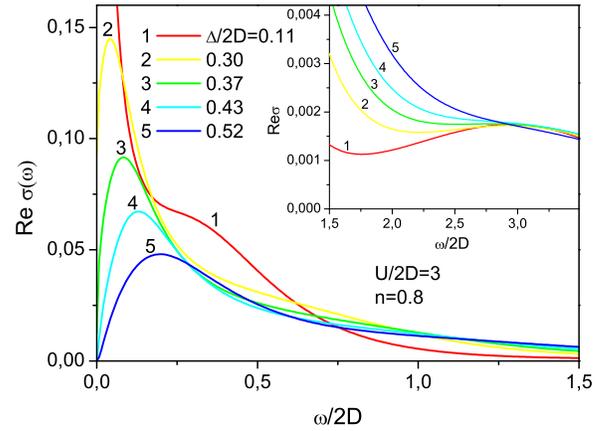}
\caption{(Color online) Real part of dynamic conductivity of Anderson--Hubbard 
model with electron concentration $n=0.8$ for different degrees of disorder 
$\Delta$ and $U=6.0D$, representing the doped Mott insulator. 
Insert -- high frequency behavior with signs of transitions to the upper 
Hubbard band.} 
\label{dop_cond} 
\end{figure}

All results presented above were obtained for half--filled case. Here we
briefly consider deviations from half--filling. In metallic phase, doping
from half--filling does not produce any qualitative changes in conductivity
behavior, which only demonstrates Anderson transition with the growth of
disorder. Thus we shall concentrate only on the case of doped Mott insulator.
Strictly speaking, for non half--filled case we never obtain Mott--Hubbard
insulator in DMFT at all. In Fig. \ref{dopDOS} we show the density of states of 
Anderson--Hubbard model with electron concentration $n=0.8$ for different 
degrees of disorder $\Delta$ and $U=6.0D$, representing typical case of the 
doped Mott insulator. Quasiparticle band overlaps now with lower Hubbard band
and is smeared by disorder, which is just the expected behavior in metallic
state. Nothing spectacular happens also with conductivity, which is shown
for the same set of parameters in Fig. \ref{dop_cond}. It shows typical
behavior associated with disorder induced Anderson MIT. Small signs of
transitions to the upper Hubbard band can be seen for $\omega\sim U$
(see insert in Fig. \ref{dop_cond}). Thus, doped Mott insulator with disorder 
is qualitatively quite similar to the case of disordered correlated metal 
discussed above. 

\section{Conclusion}
\label{concl}

We used a generalized DMFT+$\Sigma$ approach to calculate basic properties of 
disordered Hubbard model. The main advantage of our method is its ability to
provide rather simple interpolation scheme between rather well understood 
cases of strongly correlated system (DMFT and Mott--Hubbard MIT) 
and that of strongly disordered metal without Hubbard correlations, undergoing 
Anderson MIT. Apparently this interpolation scheme captures the main 
qualitative features of Anderson--Hubbard model, such as the general behavior
of DOS and dynamic (optical) conductivity. The overall picture of 
zero--temperature phase diagram is also quite reasonable and is satisfactory
agreement with the results of more elaborate numerical work \cite{BV}.
Actually, our DMFT+$\Sigma$ approach is much less time--consuming than more
direct numerical approaches, such as that of Ref. \cite{BV}, and in fact
allows us to calculate all basic (measurable) physical characteristics of
Anderson--Hubbard model.

Main shortcoming of our approach is its neglect of interference effects of
disorder scattering and Hubbard interaction, which leads to independence of
Anderson MIT critical disorder $\Delta_c$ on interaction $U$. Importance of
interference effects is known for a long time \cite{Lee85,ma}, but its 
account was only partially successful in case of weak correlations.
At the same time, the neglect of these interference effects is the major
approximation of DMFT+$\Sigma$, allowing to derive rather simple and physical
interpolation scheme, allowing the analysis of the limit of strong 
correlations. Attempts to include interference effects in our scheme are
postponed for future work.

Another simplification is, of course, our assumption of non--magnetic
(paramagnetic) ground state of Anderson--Hubbard model. The importance of
magnetic (spin) effects in strongly correlated systems is well known, as
well as the problem of competition of ground states with different types of
magnetic ordering \cite{georges96}. Importance of disorder in the studies of
interplay of these possible ground states is also quite evident. These may also 
be the subject of our future work.

Despite these shortcomings, our results seem very promising, especially
concerning the influence of strong disorder on Mott--Hubbard MIT and the
overall form of the phase diagram at zero temperature. The changes in phase
diagram at finite temperatures will be the subject of further studies.
Non--trivial predictions of our approach, such as the general behavior of
dynamic (optical) conductivity and, especially, the prediction of disorder
induced Mott insulator to metal transition can be the subject of direct
experimental verification.  

\section{Acknowledgements}

We are grateful to Th. Pruschke for providing us with his effective NRG code.
This work was supported in part by RFBR grants 05-02-16301 (MS,EK,IN),
05-02-17244 (IN), 06-02-90537 (IN), by the joint UrO-SO project (EK,IN), and programs of 
the Presidium of the Russian Academy of Sciences (RAS) ``Quantum macrophysics''
and of the Division of Physical Sciences of the RAS ``Strongly correlated
electrons in semiconductors, metals, superconductors and magnetic
materials''. I.N. acknowledges support from the Dynasty Foundation,  
International Center for Fundamental Physics in Moscow program for young
scientists and from grant of the President of Russian Federation for
young PhD MK-2118.2005.02. 

\appendix

\section{Equation for relaxation kernel}

Let us follow the standard approach of self--consistent theory of 
localization \cite{VW,WV,MS,MS86,VW92,Diagr}, taking into account the 
DMFT contributions $\Sigma^{R,A}(\varepsilon)$ in single-particle Green's
functions (\ref{GkE}) and not limiting ourselves to the usual limit of small
$\omega$. 

Consider Bethe--Salpeter equation relating the full two-particle Green's
function $\Phi^{0RA}_{\bf pp'}(\omega,{\bf q})$ to irreducible vertex
$U^{0RA}_{\bf pp'}(\omega,{\bf q})$, accounting only for impurity scattering
in vertices, but built upon Green's functions given by  
(\ref{GkE}). This equation can be written now as a generalized kinetic 
equation of the following form \cite{VW,WV,MS,MS86,VW92,Diagr}:  
\begin{eqnarray}
\left(\tilde\omega -\epsilon({\bf p})-\Delta\Sigma^{RA}_{imp}(\omega)\right)
\Phi^{0RA}_{\bf pp'}(\omega,{\bf q})=\nonumber\\
=-\Delta G_{\bf p}\left(\delta_{\bf  pp'}+
\sum_{\bf p_1}U^{0RA}_{\bf pp_1}(\omega,{\bf q})
\Phi^{0RA}_{\bf p_1 p'}(\omega,{\bf q})\right)
\label{KinEq}
\end{eqnarray}
where $\Delta G_{\bf p}=G^R(\varepsilon_+,{\bf p}_+)-G^A(\varepsilon_-,{\bf p}_-)$. 
The main difference with similar equation of 
Refs. \cite{VW,WV,MS,MS86,VW92,Diagr} is the replacement
$\omega\to\tilde\omega$. 

Let us sum both sides of (\ref{KinEq}) and of the same equation multiplied
by $({\bf {\hat p}{\hat q}})$ (where ${\bf {\hat p}}=\frac{\bf p}{|{\bf p}|}$
and ${\bf {\hat q}}=\frac{\bf q}{|{\bf q}|}$ are appropriate unit vectors)
over ${\bf p}$ and ${\bf p'}$, taking into account an exact Ward identity
\cite{VW}:  
\begin{equation} 
\Delta\Sigma^{RA}_{imp}(\omega)= \sum_{\bf p'}U^{0RA}_{\bf pp'}(\omega,{\bf q})\Delta G_{\bf p'} 
\label{TWap} 
\end{equation} 
and using an approximate representation (cf. Ref.\cite{VW})
\begin{eqnarray} 
\sum_{\bf p'}\Phi^{0RA}_{\bf pp'}(\omega,{\bf q})\approx 
\frac{\Delta G_{\bf p}}{\sum_{\bf p}\Delta G_{\bf p}} 
\Phi^{0RA}_{\varepsilon}(\omega,{\bf q})+\nonumber\\ 
+\frac{\Delta G_{\bf p}({\bf 
{\hat p}{\hat q}})} {\sum_{\bf p}\Delta G_{\bf p}({\bf {\hat p}{\hat q}})^2} 
\Phi^{0RA}_{1\varepsilon}(\omega,{\bf q})
\label{Lezhandr}
\end{eqnarray}
where $\Phi^{0RA}_{\varepsilon}(\omega,{\bf q})=
\sum_{\bf pp'}\Phi^{0RA}_{\bf pp'}(\omega,{\bf q})$ is our loop 
(\ref{PhiRA}), while $\Phi^{0RA}_{1\varepsilon}(\omega,{\bf q})=
\sum_{\bf pp'}({\bf {\hat p}{\hat q}})\Phi^{0RA}_{\bf pp'}(\omega,{\bf q})$.
Important difference from similar representation used in
Refs. \cite{VW,WV,MS,MS86,VW92,Diagr} is that (\ref{Lezhandr}) is not limited
to small $\omega$.

Now (for $q\to 0$) we obtain the following closed system of equations
defining both $\Phi^{0RA}_{\varepsilon}(\omega,{\bf q})$ 
and $\Phi^{0RA}_{1\varepsilon}(\omega,{\bf q})$:
\begin{equation}
\tilde\omega\Phi^{0RA}_{\varepsilon}(\omega,{\bf q})-
<v>q\Phi^{0RA}_{1\varepsilon}(\omega,{\bf q})=-\sum_{\bf p}\Delta G_{\bf p}
\label{SysEq1}
\end{equation}
\begin{equation}
(\tilde\omega+M(\omega ))\Phi^{0RA}_{1\varepsilon}(\omega,{\bf q})-
\frac{<v>}{d}q\Phi^{0RA}_{1\varepsilon}(\omega,{\bf q})=0\\ \nonumber
\label{SysEq2}
\end{equation}
where relaxation kernel is given by:
\begin{eqnarray}
M(\omega )=-\Delta\Sigma^{RA}_{imp}(\omega)+\nonumber\\
+d\frac{\sum_{\bf pp'}({\bf {\hat p}{\hat q}})\Delta G_{\bf p}
U^{0RA}_{\bf pp'}(\omega,{\bf q})\Delta G_{\bf p'}({\bf {\hat p'}{\hat q}})}
{\sum_{\bf p}\Delta G_{\bf p}},
\label{MKern}
\end{eqnarray}
with average velocity $<v>$ defined as:
\begin{equation}
<v>=\frac{\sum_{\bf p}|{\bf v}_{\bf p}|\Delta G_{\bf p}}
{\sum_{\bf p}\Delta G_{\bf p}};\\\ {\bf v}_{\bf p}=\frac{\partial\epsilon({\bf p})}
{\partial{\bf p}},
\label{veloc1}
\end{equation}
From (\ref{SysEq1}) we immediately obtain:
\begin{equation}
\Phi^{0RA}_{\varepsilon}({\bf q},\tilde\omega)=
\frac{-\sum_{\bf p}\Delta G_{\bf p}}
{\tilde\omega+iD(\omega)q^2} 
\label{FiRAfull} 
\end{equation}
which for small $\omega$ reduces to (\ref{FiRA}) with generalized diffusion 
coefficient given by (\ref{Dgen}). 

Using for irreducible vertex $U^{0RA}_{\bf pp'}(\omega,{\bf q})$ an 
approximation of ``maximally crossed'' diagrams and introducing the standard
self-consistency procedure of Refs.\cite{VW,WV,MS,MS86,VW92,Diagr} 
(i.e. replacing the Drude diffusion coefficient in Cooperon contribution to
irreducible vertex by the generalized one defined by (\ref{Dgen})), 
we obtain from (\ref{MKern}) our expression (\ref{MKernel}) for relaxation
kernel. 

Our equation (\ref{Dsc}) for the generalized diffusion coefficient
(which is in general case complex) reduces just to the usual transcendent 
equation. It was solved by iterations for every value of $\tilde\omega$,
taking into account that for $d=3$ and cutoff given by 
(\ref{cutoff}), the sum entering (\ref{Dsc}) reduces to:
\begin{equation}
\sum_{\bf q}\frac{1}{\tilde\omega+iD(\omega)q^2}=
\frac{1}{2\pi^2}\frac{k_0^3}{iD(\omega)k_0^2}
\int_{0}^{1}\frac{y^2dy}{y^2+\frac{\tilde\omega}{iD(\omega)k_0^2}}=
\label{Sumq}
\end{equation}
\begin{equation}
=\frac{1}{2\pi^2}\frac{k_0^3}{iD(\omega)k_0^2}
\left\{
1-{\left(\frac{\tilde\omega}{iD(\omega)k_0^2}\right)}^{\frac{1}{2}}
arctg\left({\left(\frac{iD(\omega)k_0^2}{\tilde\omega}\right)}^{\frac{1}{2}}\right)
\right\}\\ \nonumber
\label{Sumq1}
\end{equation}

For finite frequencies $\omega$ we use  
$\Phi^{0RA}_{\varepsilon}({\bf q},\tilde\omega)$ given by
(\ref{FiRAfull}), so that an expression (\ref{cond_final}) for dynamic
conductivity is to be rewritten as:
\begin{eqnarray}
{\rm{Re}}\sigma(\omega)=\frac{e^2\omega}{2\pi}
\int_{-\infty}^{\infty}d\varepsilon\left[f(\varepsilon_-)
-f(\varepsilon_+)\right]\times\nonumber\\
{\rm Re}\left\{\frac{i\sum_{\bf p}\Delta G_{\bf p}D(\omega)}
{\omega^2} -
\phi^{0RR}_{\varepsilon}(\omega)\left[1-
\frac{\Delta\Sigma^{RR}(\omega)}{\omega}\right]^2\right\},
\nonumber\\
\label{con_fin_ap}
\end{eqnarray}
The second term here was taken in the ``ladder'' approximation:
\begin{equation} 
\Phi^{0RR}_{\varepsilon}(\omega,{\bf q})=\frac{\sum_{\bf p}G^R(\varepsilon_+,
{\bf p}_+)G^R(\varepsilon_-,{\bf p}_-)}{1
-\Delta^2\sum_{\bf p}G^R(\varepsilon_+,{\bf p}_+)G^R(\varepsilon_-,{\bf p}_-)} 
\label{Ladder} 
\end{equation} 
This contribution (non singular at small $\omega$) is irrelevant for 
conductivity at small $\omega\to 0$, but leads to finite corrections 
with increasing $\omega$. Eq. (\ref{con_fin_ap}) is our final result, which
was analyzed numerically in a wide interval of frequencies (for small 
$\omega$ it reduces to (\ref{con_fin})).

\section{``Bare'' electron dispersion and velocity}

We consider the ``bare'' energy band with semi-elliptic DOS (\ref{DOS}). 
Assuming isotropic electron spectrum $\epsilon ({\bf p})=\epsilon (|{\bf p}|)
\equiv\epsilon (p)$ and equating the number of states in spherical layer of
momentum space to the number of states in an energy interval 
$[\epsilon ,\epsilon +d\epsilon ]$], we obtain differential equation
determining the energy dispersion $\epsilon (p)$:
\begin{equation}
\frac{4\pi p^2dp}{(2\pi)^3}=N_{0}(\epsilon )d\epsilon
\label{DifEq1}
\end{equation}
For quadratic energy dispersion $\epsilon (p)$ close to the lower band edge
we get the initial condition for Eq. (\ref{DifEq1}) for $p\to 0$ and 
$\epsilon\to -D$. Then we obtain
\begin{equation}
p={\left[6\pi\left(\pi-\varphi +\frac{1}{2}sin(2\varphi )\right)
\right]}^{\frac{1}{3}}
\label{spektr1}
\end{equation}
with $\varphi =arccos(\frac{\epsilon}{D})$ and momentum in units of inverse
lattice spacing. Eq. (\ref{spektr1}) implicitly defines ``bare'' energy 
dispersion $\epsilon (p)$ for electronic part of the spectrum 
$\epsilon\in [-D,0]$.

For half-filled band we easily determine the Fermi momentum as:
\begin{equation}
p_{F}=p(\epsilon =0)={\left(3\pi^2\right)}^{\frac{1}{3}}
\label{pF}
\end{equation}
We also need electron velocity $|{\bf v_p}|=
\left|\frac{\partial\epsilon({\bf p})}{\partial{\bf p}}\right|=
\frac{\partial\epsilon(p)}{\partial p}$ entering into the expression 
(\ref{veloc1}) for average velocity.
From (\ref{DifEq1}) we obtain:
\begin{equation}
|{\bf v_p}|=\frac{d\epsilon}{dp}=\frac{p^2}{2\pi^2}\frac{1}{N_0(\epsilon )}
\label{veloc_e1}
\end{equation}
where $p$ is given by Eq. (\ref{spektr1}).

To obtain quadratic dispersion for hole part of the spectrum 
($\epsilon\in [0,D]$) close to the upper band edge ($\epsilon\to D$) we 
introduce the hole momentum $\tilde p=2p_{F}-p$  and write similarly to
(\ref{DifEq1}):
\begin{equation}
\frac{4\pi\tilde p^2d\tilde p}{(2\pi)^3}=-N_{0}(\epsilon )d\epsilon
\label{DifEq2}
\end{equation}
Putting $\tilde p\to 0$ at the upper band edge $\epsilon\to 0$, we get:
\begin{equation}
\tilde p={\left[6\pi\left(\varphi -\frac{1}{2}sin(2\varphi )
\right)\right]}^{\frac{1}{3}}
\label{spektr2}
\end{equation}
Then for velocity at the hole part of the spectrum we obtain:
\begin{equation}
|{\bf v_p}|=\frac{d\epsilon}{dp}=-\frac{d\epsilon}{d\tilde p}=
\frac{\tilde p^2}{2\pi^2}\frac{1}{N_0(\epsilon )}
\label{veloc_e2}
\end{equation}
Eqs. (\ref{veloc_e1}) and (\ref{veloc_e2}) define energy dependence of 
$|{\bf  v_p}|$. It is easily seen that velocity is even in energy and becomes
zero at the band edges. These expressions allow us to change from momentum
summation (e.g. in Eq. (\ref{veloc1})) to energy integration.

\pagestyle{empty}

\newpage

\end{document}